# The Frascati Beam Test Facility

B. Buonomo, F. Cardelli, C. Di Giulio*, D. Di Giovenale, L. G. Foggetta,

C. Taruggi, INFN-LNF, [00044] Frascati, Italy

*Abstract*— From 2004 the Frascati Beam Test Facility (BTF) in the DAΦNE accelerator complex provides to the external user up to $10^{10}$ electrons per bunch or up to $10^9$ positrons per bunch to develop their detector.

After an upgrade program terminated in 2020 of the beam test facility a description of the status and available beam lines will be done.

*Keywords—Beamline, LINAC, Detectors,*

## I. Introduction

The test beam and irradiation facilities are critical enabling infrastructures in the development of detectors for high energy physics (HEP) experiments such as the Future Collider and astroparticles physics experiments.

In the European infrastructures for the development and testing of particle detectors, the Beam-Test Facility (BTF) of the DAΦNE accelerator complex at the Frascati laboratory of the Italian National Institute of Nuclear Physics (INFN) has assumed a significant role since 2004. [1–7].

In 2016 a proposal was written by the BTF team with the intention of enhancing the facility's functionality and expanding the LINAC beam's potential applications to the BTF lines, with the goal of hosting long-term investigations in basic physics [4] and offering electron irradiation to industrial users as in ERAD project.

The respond at those requirements the BTF beamline needed to be doubled to handle users and at the same time to maintain the long-term installation as PADME in one of the beamlines [3].

The original BTF line has been operating in a spare pulse mode when the DAΦNE electron-positron collider is operational [10, 11]. Within the stated radio-protection criteria for the current shielding configuration, the entire Frascati LINear Accelerator (LINAC) beam could also be extracted in the direction of the BTF line without being conditioned by the target as described in section III.

In the next paragraphs the description of the LINAC and of the beam line measured characteristics respect the Conceptual Design Report [12] will be discussed to provide at the reader the information to understand if the BTF will be compliant at their requirements for their test beam and how to apply for beam time request

## II. The LINAC

### A. The gun and RF distribution

The Frascati Linear Accelerator is used as the electron and positron source for the *DAΦNE* [1] Collider and Beam Test Facility (BTF) [2], and for experiments with fixed targets such as PADME [3] and irradiation tests of space components. After commissioned in 1996, the modernization of the 120KV electrostatic gun pulser as allow to be compliant at the requirement for a bunch length bigger and smaller the designed 10 ns [8] and an upgrade on the conventional L-C resonant charging system for the RF plant was start in 2018 allow to be compliant to the energy stability. Currently, three out of four RF power plant modulators are being upgraded from three-phase variable phase control (SCR) based full-wave bridge diode assemblies to new constant current capacitor charging power supplies. The LINAC 15 accelerating structures are powered by four high power S-band pulsed klystrons (45 MW model TH2128 C) driven by Pulse modulators deliver high voltage pulses of 315 kV, 360 A with a pulse duration of 4.5 μs and a nominal pulse repetition frequency of 50 Hz. The RF power distribution in Fig.1 shown where the positron converted, and the positron separator are placed along the LINAC. The RF power by the SLED is distributed by the waveguide network to fill the accelerating structures [9].

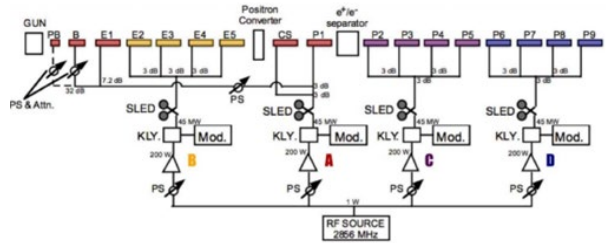

Figure 1 The LINAC RF distribution.

### B. The LINAC beam charateristics

| | Design | Operational (top) |
|---|---|---|
| Electron beam final energy | 800 MeV | 510 MeV (750) |
| Positron conversion energy | 250 MeV | 220 MeV |
| Positron beam final energy | 550 MeV | 510 MeV (535) |
| RF frequency | 2856 MHz | |
| Accelerating structure | SLAC-type, CG, 2π/3 | |
| RF Amplifiers | 4 x 45 MW sledded klystrons TH2128C | |
| Beam pulse rep. rate | 1 to 50 Hz | 1 to 50 Hz |
| Beam macropulse length | 10 nsec | 1.4 ns to 300 ns |
| Beam spot on positron converter | 1 mm | 1 mm |
| Normalized Emittance (mm mrad) | 1 (electron) 10 (positron) | 1 (electron) 10 (positron) |
| RMS Energy spread | 0.5% (electron) 1.0% (positron) | 0.5% (electron) 1.0% (positron) |
| Output electron current (510MeV) | >150 mA | 180 mA (>500) |
| Output positron current (510MeV) | 36 mA | 50 mA (>85) |

Table 1 Frascati LINAC parameters

In Table 1 the LINAC design parameters and the operative ones are summarized.

The energy spread of the beam of 510 MeV provided at the *DAΦNE* collider for electron and positron are shown in Fig. 2 and 3. This was measured by the spectrometer named as DHSTS001 in figure 5. In Fig. 4 the measured beam for the ERAD project in the BTF was shown for the 165 MeV beam with a charge of 400 pC required. To obtain this beam three of the 4 RF plant was in accelerating mode and the last RF plant the provides the RF power to the last four accelerating sections was in decelerating mode to conserve the beam quality.

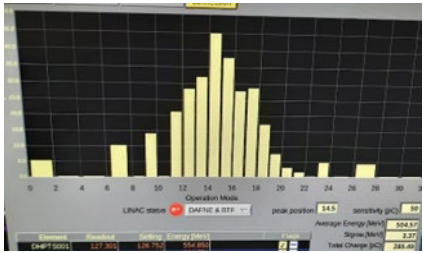

Figure 2 Spectrometer image of LINAC positron beam with an energy of 510 MeV and 250pC of charge.

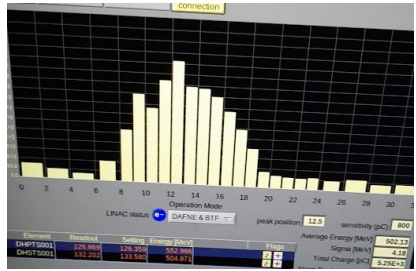

Figure 3 Spectrometer image of LINAC electorn beam with an energy of 510 MeV and 5.8 nC of charge.

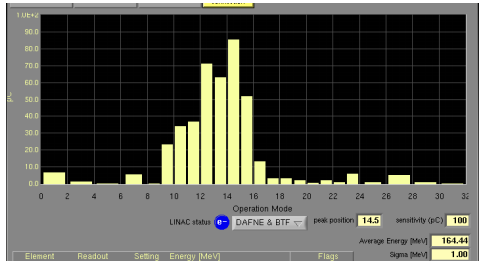

Figure 4 Spectrometer image of LINAC positron beam with an energy of 164 MeV and 400pC of charge.

## III. THE BTF BEAMLINES

### A. The upgrade of the BTF

At the Frascati Beam-Test Facility (BTF), the primary requirements from users for detector testing and beam-test activities can be succinctly outlined as follows:

- Optimal Beam Quality: users demand a high-quality beam, particularly focusing on beam size, divergence, and background, even at the low end of the BTF energy range (few tens of MeV). Accommodating this need becomes particularly challenging for setups positioned in air downstream of the exit window.

- Extending the Energy Range: researchers aim to extend the energy range of their detectors from tenths of MeV up to maximum energy to understand the response of the detector from single particle to maximum number of particles.

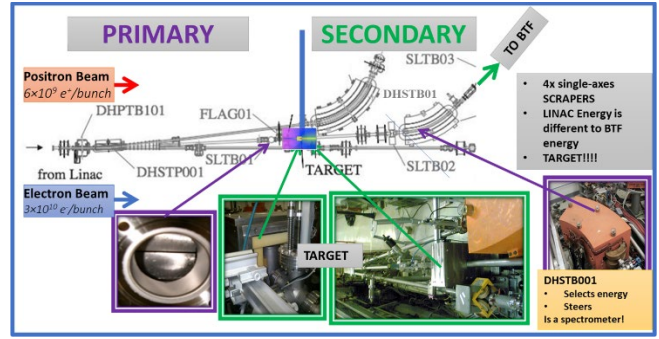

Figure 5 The BTF secondary beam production system

The Beam-Test Facility (BTF) serves as an extraction and transport line, generating electrons or positrons across an extensive range of intensity, energy, beam spot dimensions, and divergence. It originates from the primary beam of the LINAC, which accelerates 50 pulses/s. One pulse is directed to the spectrometer, while the other can be directed to a small ring for emittance damping (subsequently injected into the collider rings) or to the BTF line using pulsed dipoles.

To broaden the momentum distribution of the incoming beam, a variable depth target (from 1.7 to 2.3 X0) is employed. Following this, secondary electrons (or positrons) undergo momentum selection using a 45-degree bending dipole and collimators in the horizontal plane. As a result, the beam intensity is significantly reduced, contingent upon the chosen secondary beam energy central value (ranging from about 25 MeV to nearly the primary beam energy) and spread (typically better than 1 percent at higher energy, depending on collimator settings).

The transported beam is then directed to the first experimental hall and focused using two quadrupole FODO doublets. For clarity, refer to Figure 1, which illustrates the layout of the beam selection and transport line, along with the shielded experimental area.

In terms of recent developments, the new layout concept includes a beam-splitting dipole wrapped around a double-exit pipe. This innovative design allows for alternating beam pulses from the upstream BTF beamline between two new lines [9]. The dipole is connected to a pulsed magnet power supply and could enable the switching between the two lines, but for the internal rules we use only one beamline per time.

The first line is dedicated to directing the beam to the existing experimental hall ("BTF EH1"), utilizing the pre-existing concrete blockhouse. Meanwhile, the second line transports the beam, with the addition of three extra dipoles, to the area previously utilized as the BTF control room ("BTF EH2").

To achieve optimal performance, a comprehensive optimization of the new lines' optics has been conducted. Defining the requirements for the new beam elements involved using advanced simulation tools such as G4beamline and MADX [13,14]. After evaluating various options, the original notion of employing a fast magnet for a 45-degree bending angle was abandoned in favor of a fast 15-degree dipole magnet.

Due to the momentum dispersion introduced by the bending magnet, the relative energy spread ΔE/E is essentially determined by the magnet/collimator configuration [10,11]; in the standard BTF operation for a wide range of slit apertures a resolution better than 1% can be obtained.

The number of transported electrons (or positrons) can be adjusted in a wide range, down to single particle, and is well below the sensitivity of any standard beam diagnostics device, so that many different particle detectors have been used to monitor the beam characteristics.

In the following sections a description of the two lines available in BTF1 and the new line in BTF2 will be discussed.

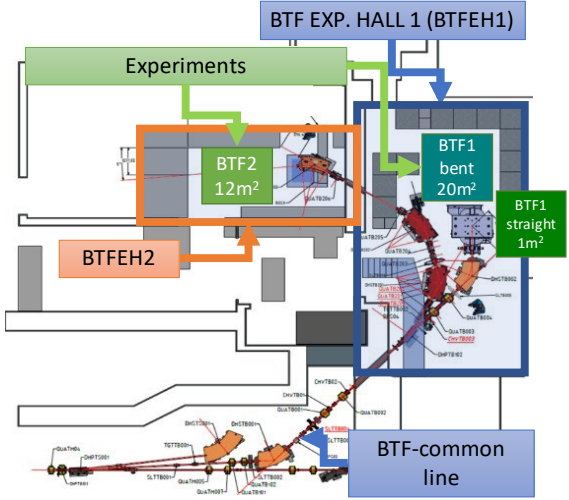

Figure 6 The BTF transfer line and experimental area.

In the BTF1 the beam features are exposed in table 2 for a dedicate mode and in time sharing mode with DAΦNE.

| Parameters | BTF1 Time sharing | | BTF1 Dedicated | |
|---|---|---|---|---|
| | With Cu target | Without Cu target | With Cu target | Without Cu target |
| Particle | $e^+ / e^-$ (User ) | $e^-$ (DAΦNE status) | $e^+ / e^-$ (User ) | $e^-$ |
| Energy (MeV) | 25–500 | 510 | 25–700 ($e^-/e^+$) | 167–700 ($e^+$) 250–550 ($e^-$) |
| Best Energy Resolution at the experiment | 0.5% at 500 MeV | 0.5%/1% | 0.5%(Energy/mult dependent) | |
| Repetition rate (Hz) | Variable from 1 to 49 (DAΦNE status) | | 1–49 (User) | |
| Pulse length (ns) | 10 | | 1.5–320 (User) | |
| Intensity (particle/bunch) | 1–10⁵ (Energy dependent) | $10^9$ to $1x10^{10}$ | 1–10⁵ (Energy dependent) | 1 to 3x10¹⁰ |
| Max int flux | 3.125x10¹⁰ part./s | | | |

Table 2 BTF1 beam parameters.

| Parameters | BTF2 Time sharing | BTF2 Dedicated |
|---|---|---|
| | With Cu target | With Cu target |
| Particle | $e^+ / e^-$ (User ) | |
| Energy (MeV) | 25–500 | 25–700 |
| Best Energy Resolution at the experiment | 1% at 500 MeV (energy/mult. dependent) | |
| Repetition rate (Hz) | Variable from 1 to 49 (DAΦNE status) | 1–49 (User) |
| Pulse length (ns) | 10 | 10 |
| Intensity (particle/bunch) | 1–10⁴ (energy dependent) | |
| Max int flux | 1x10⁶ part./s | |

Table 3 BTF2 beam parameters.

In table 3 the BTF2 beam parameters are summarized, for the limitation on the radioprotection up to $10^6$ particles per second the beams is always conditioned by the copper target described before.

### B. The BTF1 Straight beamline

The Experimental Hall 1 Straight Line is an experimental area of 1m² with a remote controlled table shown in figure 8. It is dedicated to the high intensity internal test beam usually in the BTF dedicated mode.

On this line a OTR acquisition system and cameras for YAG flags too it is implemented. Those has provided the possibility for staff to monitor the beam parameters and to provide a first measurement of the vertical emittance for the positron beam of the LINAC measured with a quadrupole scan made by the magnetic elements shown in figure 9 in dedicated mode. The positron beam size at waist measured with the setup shown in figure 10 and the emittance in function of the quadrupole current is shown in figure 11, where positron beam vertical emittance measured was 0.93 ± 0.2 mm mrad for the energy beam of 497 MeV for a bunch length of 10 ns and 4.7 pC of charge.

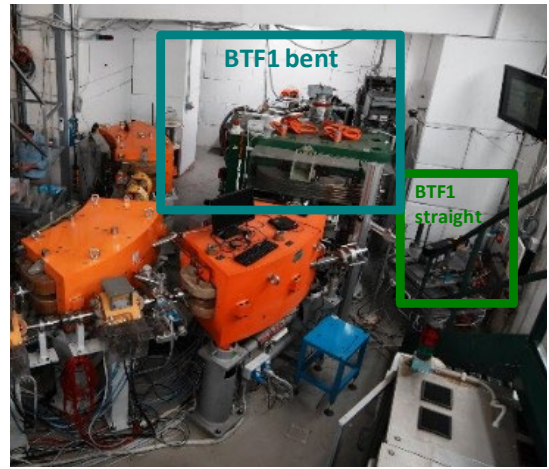

Figure 7 The exmpaerimental areas in BTF1 experimental hall.

This line, used to check the beam spot for PADME experiment, was characterized for the operation with the CU target for low energy beam and a collection of the measured beam size ad different energy obtained with a silicon pixel detector with a sensor dimension of 1,7 x 1,7 cm was shown in figure 12.

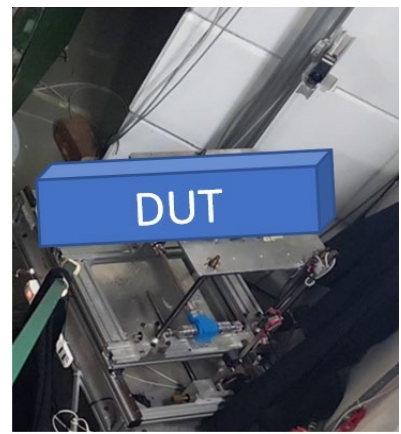

Figure 8 BTF1 remote controlled table.

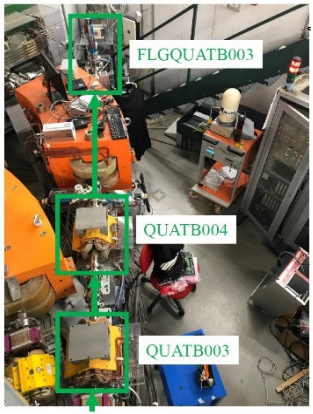

Figure 9 The BTF1 straght line magnetic elements, the OTR screen mechanical movmeent and the quadrupoles for the measurmeents are marked in green line.

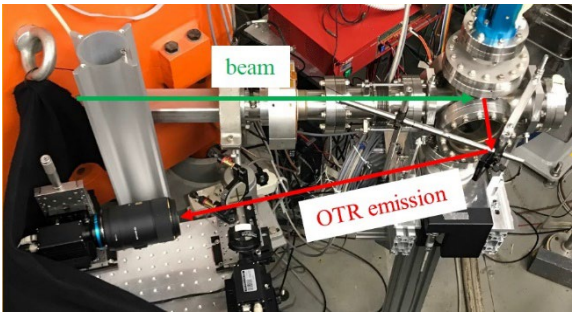

Figure 10 the BTF1 OTR optical path for emittance mesurement.

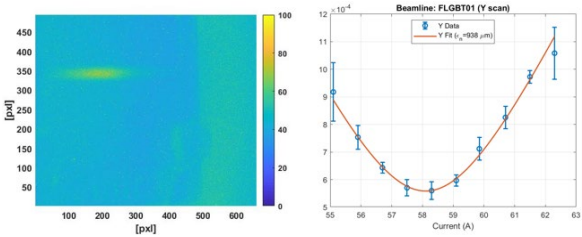

Figure 11 on the left the positron beam spot size at the waist, on the right the quadrupole scan for vertical emittance mesurement.

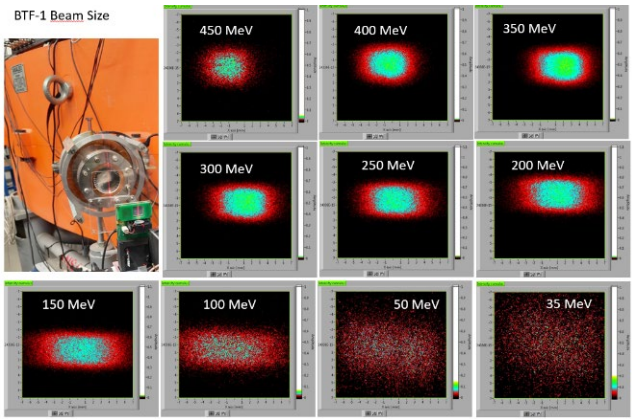

Figure 12 BTF1 secondary beam spot size at different energy.

## C. The BTF1 Bent line

The Experimental Hall 2 Bent Line is an experimental area of 20m² dedicated to long term experiment.

The area is currently occupied by PADME (Positron Annihilation into Dark Matter Experiment) [12,13]. The experiment is searching for the possible mediator of a new interaction between Standard Model particles and dark matter particles. PADME studies the interaction between a positron beam (nominal energy 550 MeV) and an active diamond target. The annihilation of the positrons with the target electrons could produce the dark photon A', according to the process e⁺e⁻ → A'γ. After the interaction, the charged particles that did not interact with the target are deflected in vacuum by a magnetic field towards the charged particle veto detectors, while the neutral particles reach the electromagnetic calorimeter of the experiment (ECal). ECal is a segmented BGO calorimeter, in charge of the detection of the SM photon in the final state. The Bremsstrahlung radiation, which is one of the main backgrounds of the experiment and emitted with a narrow angle with respect to the beam direction, is detected by the small angle calorimeter, placed behind the central hole in the ECal. If the ECal detects the photon, the kinematics of the process can be closed, and the existence of the dark photon could be identified as a peak in the missing mass distribution of the process: in particular, with a 550 MeV, the peak should correspond to a dark photon mass of 23.7 MeV. The beam request for the experiment was spot size of 0.5 mm x 1 mm with a bunch length of 300 ns from 30000 positron down to 3000 positron that was characterized by the BTF silicon pixel detector and verified by the diamond detector for the spot size [4] with a particle distribution along the 300 ns verified by the PADME calorimeter [3] as shown in Fig. 15 and 16.

During 2022, the detector has been modified to perform a new measurement, the detection of the X17 boson. This particle could be responsible for the anomalous decay of beryllium observed by the A. Krasznahorkay group from ATOMKI [14]. In this case, the beam energy was not constant, but an energy scan was performed. BTF was the only facility in the world that was able to deliver a positron beam with the precise characteristics to perform the resonant production of the X17 boson. One critical parameter for the signal to background optimization of the experiment was the beam energy spread [15].

The area is occupied by the PADME experiment with the aims was the research of dark matter candidates as shown in Fig. 13. The experiment works thanks to a small but extremely precise measuring apparatus, able to observe the production of the dark photon in electron-positron collisions. PADME is installed in the experimental hall of the test facility (BTF) of the linear accelerator of the INFN Frascati National Laboratories. Accelerated positrons hit on to a thin diamond target. By interacting with the atomic electrons, positrons could produce "dark photons" together with a visible photon.

The positrons that interact with the target are then deflected away from the detectors with/through the magnetic field produced by a dipole.

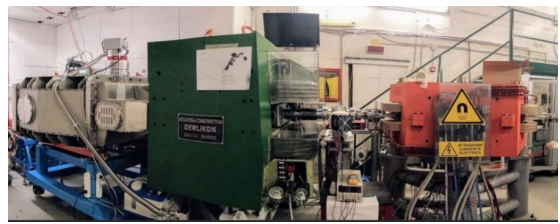

Figure 13 PADME experiment installed in BTF1 bent line.

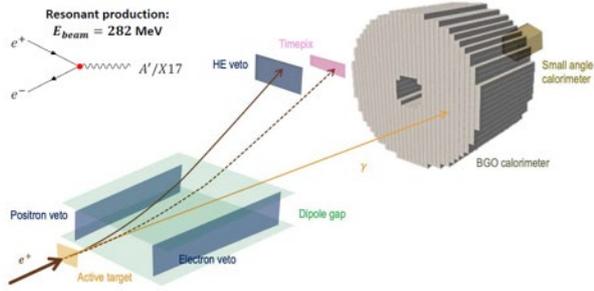

Figure 14 The PADME experiment concept.

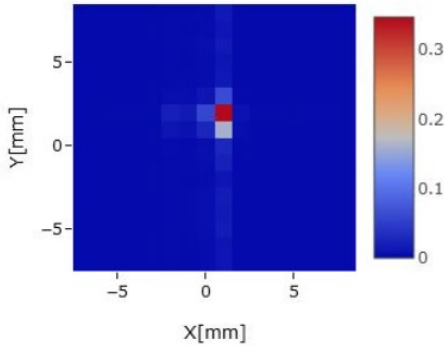

Figure 15 PADME beam spot size measured with diamond detector.

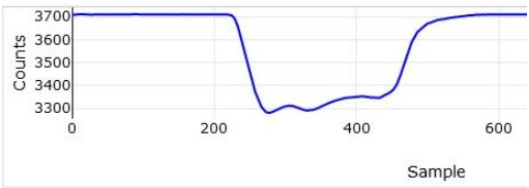

Figure 16 Poistron distribution along time for a 200 ns bunch lenght mesured by PADME calorimeter.

### D. The BTF2 line

The BTF line 2 has an experimental area of 12 m² and a remote controllable table for the users as shown in figure 17. In this area different diagnostic is installed for the low intensity beam available for this area [17-26]. The silicon pixel detector (FitPix and TimePix3) and lead glass calorimeter are available for the user to control the beam quality during the test beam as illustrated in figure 18.

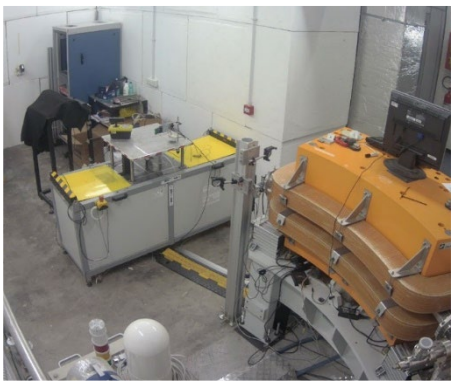

Figure 17 The BTF experimental area. The diagnostic and the remotable table are visible.

Due to the good energy resolution of the calorimeter, the number of produced electrons can be counted simply by measuring the total deposited energy E: n = E/E1, where E1 is the energy deposited by a single electron. An example of ADC spectrum is shown in Fig. 18, for a selected energy of $E_{sel}$ = 450 MeV the individual peaks corresponding to 0, 1, . . . , n electrons can be easily detected. The total number of events in each peak should represent the probability of producing n particles, by fitting the distribution of the number of events in each peak with the Poisson function. the average number of particles can be determined.

The spot size obtained with the silicon pixel detector at the Aluminium exit windows of the beamline is shown in figure 19 for different secondary beam energy.

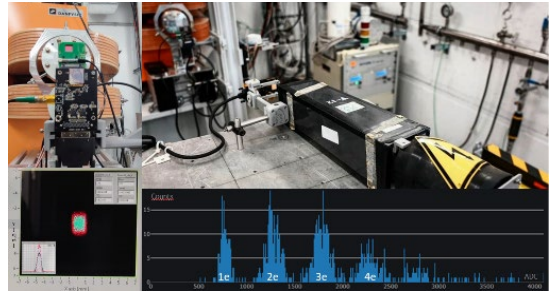

Figure 18 The BTF2 beam diagnostics [16-22].

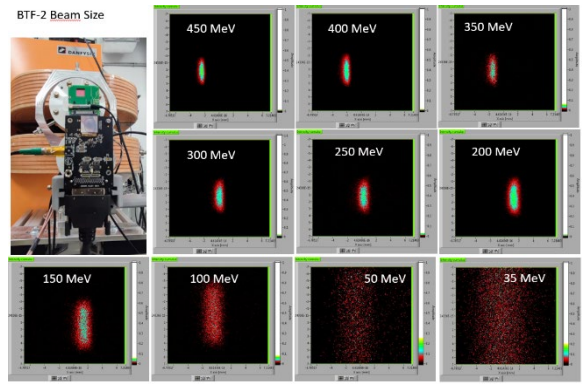

Figure 19 BTF2 secondary beam size for different energy mesured by a silicon pixel detector.

## IV. CONCLUSIONS

The Frascati LINAC modulator Upgrade is almost concluded.

The Beam Test Facility of Laboratori Nazionali di Frascati provides e⁺/e⁻ beam in different configurations for particle detectors developers. BTF can deliver to user e⁺/e⁻ beam with multiplicity ranging from 1 particle/bunch to $10^{10}$ particles/bunch, and energy up to 700 MeV (e⁻) and 500 MeV (e⁺).

The facility consists of two experimental hall and three beamlines, two of the three available beamlines are equipped with remote-controlled tables and beam diagnostics.

A call for users opens twice a year and users are selected by a scientific committee and could request 1 or 2 weeks of beam time.

The beam time could be requested to https://btf.lnf.infn.it/schedule-beam-request/ .

ACKNOWLEDGMENT

We would like to thank to LNF Staff that support the BTF with their expertise in safety, administrative and technical issue supporting the call for proposal to test beam.